# A Machine Learning Approach for Delineating Similar Sound Symptoms of Respiratory Conditions on a Smartphone

Chinazunwa Uwaoma[1] and Gunjan Mansingh[1]

[1] Department of Computing, The University of West Indies, Kingston 7, Jamaica

`chinazunwa.uwaoma@mymona.uwi.edu, gunjan.mansingh@uwimona.edu.jm`

**Abstract.** Clinical characterization and interpretation of respiratory sound symptoms have remained a challenge due to the similarities in the audio properties that manifest during auscultation in medical diagnosis. The misinterpretation and conflation of these sounds coupled with the comorbidity cases of the associated ailments – particularly, exercised-induced respiratory conditions; result in the under-diagnosis and undertreatment of the conditions. Though several studies have proposed computerized systems for objective classification and evaluation of these sounds, most of the algorithms run on desktop and backend systems. In this study, we leverage the improved computational and storage capabilities of modern smartphones to distinguish the respiratory sound symptoms using machine learning algorithms namely: Random Forest (RF), Support Vector Machine (SVM), and k-Nearest Neighbour (k-NN). The appreciable performance of these classifiers on a mobile phone shows smartphone as an alternate tool for recognition and discrimination of respiratory symptoms in real-time scenarios. Further, the objective clinical data provided by the machine learning process could aid physicians in the screening and treatment of a patient during ambulatory care where specialized medical devices may not be readily available.

## 1 Introduction

Respiratory sounds such as cough, sneeze, wheeze, stridor, and throat clearing are observed as clinical indicators containing valuable information about common respiratory ailments. Conditions such as Asthma, Vocal Cord Dysfunction (VCD), and Rhinitis provoked by prolonged and vigorous exercise, are often associated with these symptoms which sometimes overlap; thus, making it difficult for proper diagnosis and treatment of the underlying ailment symptomized by the respiratory sounds. Given the similarity of their acoustic properties, these sounds at times, are conflated and misinterpreted in medical assessment of patients with respiratory conditions using conventional methods. Further, the evaluation of these sounds is somewhat subjective to physicians' experience and interpretation, as well as the performance of the medical device used for monitoring and measurement [1], [2].

To address these issues, several studies in recent times have proposed different approaches for objective detection and classification of respiratory sounds using computerized systems. However, with improvement on the storage and computational capabilities of mobile devices, there is a gradual move from the use of specialized medical devices and computer systems to wearable devices for recording and analysing respiratory sounds in real-time situations [3], [4]. Much efforts have been focused on the analysis of wheezing sound given its clinical importance in the evaluation of asthma, COPD and other pulmonary disorders [5]. Considerable attention has also been given to physiological mechanism and formation of other pathological respiratory sounds such as stridor, cough, and crackles [3], [6]. At times, these sounds appear together on the same respiratory signal and their accurate detection and classification remain subjects of interest to many researchers [7], [8], [9].

The sound symptoms specifically, bronchial asthma wheezes and VCD stridor are often confused in the preliminary diagnosis of airways obstruction during physical exercise [10]. Both sounds have been described as continuous, high-pitched musical sounds. They also exhibit periodicity in time domain given their sinusoidal waveforms. However, stridor is said to be louder and can be heard around the neck without the aid of a stethoscope. Dominant frequencies are between 100 - 1000Hz [6]. Wheeze on the other hand, originates from the bronchia and it is mostly audible around the chest wall [11], with dominant frequencies around 600Hz [12]. Other respiratory sounds heard in the events of air passage obstruction or irritation include cough, throat clearing, sneezing and sniffle. Unlike wheeze and stridor, these categories of sounds are percussive, transient, and have quasi-periodic wave forms and short duration. Apart from audio information of the symptoms, there are other factors used in the differential diagnosis of exercised-induced asthma and VCD such as the respiratory phase of the sound occurrence (Inspiratory/Expiratory/Biphasic), and the reversibility of conditions [6], [10], [11]. However, these issues are not within the scope of this paper.

The study objective is to distinguish acoustic properties of respiratory symptoms that correlate with certain respiratory conditions induced by highly intensive physical activity; using smartphone as a platform for the analysis and classification of the sounds. The approach focuses on time-domain and frequency-domain analysis of these sounds. The machine learning algorithms exploit the differences in the energy content and variation, periodicity, spectral texture and shape as well as localized spectral changes in the signal frames. The extracted features from the audio data analysis are fed into classifiers - Random Forest, support vector machine (SVM), and k-Nearest Neighbor (k-NN). The classification algorithms are performed on both individual domain and combined domain feature sets. A leave-one-out approach is used in the evaluation of the performance of the classifiers for objective comparison of their discriminatory abilities.

## 2 Related Work

Recent studies have focused on audio-based systems for continuous monitoring and detection of vital signs relating to management and control of long-term respiratory conditions. Aydore et al. in their work [1], performed a detailed experiment on the classification of wheeze and non-wheeze episodes in a respiratory sound, using linear analysis. Though the approach they adopted yielded an impressive success rate of 93.5% in the testing; the study was not specific about the non-wheeze category of sounds such as rhonchi and stridor which mimic wheeze, and are reportedly misdiagnosed as wheeze in clinical practice. The work however, was extended by Ulukaya et al. [7] on the discrimination of monophonic and polyphonic wheezes using time-frequency analysis based on two features – mean crossing irregularity (MCI) in the time domain, and percentile frequency ratios in the frequency domain. The authors considered MCI as the best discriminating feature with a performance accuracy of 75.78% when combined with image processing.

There are on-going research efforts towards the design of monitoring and detection systems for respiratory conditions based on mobile platforms. The overall aim of these studies is to increase the awareness and compliance by individuals in managing their conditions, and to improve the efficacy of treatment procedures and therapies by health professionals. In the study [3], mobile phone was used as a sensing platform to track cough frequency in individuals and across geographical locations. The embedded microphone in the mobile phone serves as audio sensor to record cough events, with the phone placed in the shirt or pant pockets or strapped on the neck of the user. According to the authors, results obtained from the study could be used in further diagnosis and treatment of diseases such as pneumonia, COPD, asthma, and cystic fibrosis. Automated Device for asthma Monitoring (ADAM) was developed by Sterling et al. [13], to monitor asthma symptoms in teenagers. The system design involves the use of lapel microphone attached to the mobile and worn by the user to capture audio signals. It uses Mel-frequency cepstral coefficients (MFCC) and multiple Hidden Markov Model (HMM) for feature extraction and classification, to detect the 'presence' or 'absence' of cough in the recorded sounds. The sensitivity of the detection algorithm is 85.7%. BodyBeat, proposed by Rahman et al. [14], is another mobile sensing system for recognition of non-speech body sounds. Like ADAM, it uses a custom-made microphone attached to an embedded unit (Micro-controller) for audio capturing and pre-processing. The embedded unit connects to the mobile phone through Bluetooth for feature extraction and classification of the audio windows. Sun et al. [15] in their study, proposed SymDetector, a mobile application for detection of acoustic respiratory symptoms. The application samples audio data using smartphone's built-in microphone and performs symptom detection and classification using multi-level coarse classifier and SVM.

Though the designs appear quite elaborate and plausible on the mobile platform; common issues with these approaches include the ease of use of the system, and the reproducibility of the algorithms used in the detection process. There could be concerns about the setup and cost of deployment by the user for systems that utilize external audio sensors and other devices connected to the mobile phone. Also, running multiple level classification for the detection algorithms may impact on the response time of the applications when deployed in real-time. In addressing these issues, our study uses a standalone mobile platform with no external gadgets connected to the smartphone. This allows all the major operations – audio sampling, pre-processing, feature extraction, and classification to be performed on the mobile phone.

The next section of the paper describes the methods used in audio data acquisition, pre-processing and analysis techniques, and feature extraction. Section 4 highlights the classification algorithms and feature sets for the classifiers. In section 5, the classification results and performance evaluation are discussed. Section 6 describes application scenarios of the classification results while the conclusion on the study is provided in the last section.

## 3 Methods

### 3.1 Sound Recordings and Datasets

To ensure the reliability and fidelity of the datasets, the recordings used in this study were retrieved from different but trusted sources. The wheeze and stridor sounds are specifically collected under licensed agreement, from R.A.L.E Lung repository [16]; with each record, pre-labelled by an expert physician. The cough, throat clearing, and other sounds are obtained from direct recordings from healthy individuals and pathological subjects using the mobile phone microphone. The dataset comprises of five categories of sound including: wheeze, stridor, cough, throat clearing, and a mixed collection of other sounds. By visual inspection of the waveforms and audio verification, all distinct segments of the audio recordings containing the actual sounds are selected. Given the varying length and sampling rate of the recordings obtained from the repository, the audios are down-sampled to 8000Hz and segmented into equal length to ensure uniformity and to lessen computational load on the mobile device.

### 3.2 Signal Pre-Processing and Analysis

The techniques used in the signal pre-processing and analysis include windowing and digitization of each audio signal into frames of equal length (128ms) with 87.5% overlap. The signal frames are decomposed into spectral components using the Discrete Short-Time Fourier Transform (STFT) technique. Hamming window of size N = 1024 was used to reduce spectral leakage in the signal frames. The windowing and overlapping techniques help to smoothen the spectral parameters that vary with time.

### 3.3    Feature Extraction

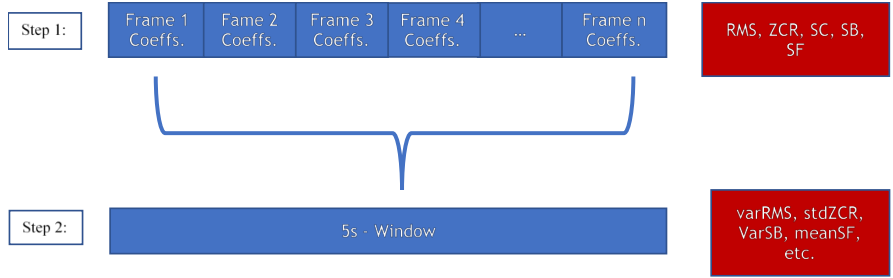

**Fig. 1.** A two-level step for Feature Extraction.

In preparing the feature sets for classification, we employed two steps in the feature extraction. First, is the frame-level extraction, where the resulting coefficients from STFT served as input parameters for calculating the temporal and spectral instantaneous features of the audio signals. The window-level features or texture features are derived from the instantaneous features. These features are basically statistical functions of the frame-level features expressed in terms of rate of change, extremes, averages, and moments of grouped frames - about 5 seconds of the audio duration as shown in Fig. 1.

Time-domain features used include the RMS energy and Zero Crossing Rate (ZCR) of each frame in the audio record. Some of spectral features used in the classification are described as follows:

**Spectral Centroid (SC).** This feature measures the spectral shape of individual frames and it is defined as the centre of spectral energy (power spectrum). Higher values indicate "brighter" or "sharper" textures with significant high frequencies, while lower values correspond to low brightness and much lower frequencies. Given $P$ as the power spectrum of the frame $f$, and $N$ being the Nyquist frequency with $k$ as the frequency bins; SC is calculated as:

$$SC(f) = \frac{\sum_{k=0}^{N-1} k . P_k^2}{\sum_{k=0}^{N-1} P_k^2} \qquad [17] \qquad (1)$$

**Spectral Bandwidth (SB)**. Also known as 'instantaneous bandwidth' [18], SB technically describes the spread or concentration of power spectrum around the SC. It is a measure of 'flatness' of the spectral shape. Higher values often indicate noisiness in the input signal and hence, wider distribution of the spectral energy; while low values show higher concentration of the spectral energy at a fixed frequency region. SB is calculated as follows:

$$SB(f) = \sqrt{\frac{\sum_{k=0}^{N-1}(k - SC(f))^2 . |P_k|^2}{\sum_{k=0}^{N-1} P_k^2}} \qquad [17] \qquad (2)$$

**Spectral Flux(SF)**. Spectral Flux is an approximate measure of the sensation 'roughness' of a signal frame [18]. It is used to determine the local variation or distortion of the spectral shape and it is given by:

$$SF(f) = \frac{\sqrt{\sum_{k=0}^{N-1}(|P_f| - |P_{f-1}|)^2}}{N-1} \qquad [17] \qquad (3)$$

Table 1 provides a full list of the frame-level and window-level features.

**Table 1.** Classification Features [17].

| | Feature Group | Descriptor | Classification Acronym |
|---|---|---|---|
| **Frame Level** | Energy | Root Mean Square | RMS |
| | Periodicity | Zero Crossing Rate | ZCR |
| | Spectral Shape | Spectral Centroid | SC |
| | | Spectral Bandwidth | SB |
| | | Spectral Flux | SF |
| **Window Level** | Extremes | AMR of RMS window [15] | amrRMS |
| | | Relative Max RMS [15] | rmrRMS |
| | Averages | Mean of RMS window | meanRMS |
| | | Mean of SC window | meanSC |
| | | Mean of SB window | meanSB |
| | | Mean of SF window | meanSF |
| | Moments | Variance of RMS window | varRMS |
| | | Std. of ZCR window | stdZCR |
| | | Mean Crossing Irregularity [7] | mciZCR |
| | | Variance of SC window | varSC |
| | | Variance of SB window | varSB |
| | | Variance of SF window | varSF |

# 4 Classification Algorithms

In the study, three classifiers – Random Forest, kNN, and SVM were used to investigate the performance of the extracted input parameters in differentiating the audio sound patterns. Each of the classifiers represents a category of classification algorithms often used in Machine Learning. Whereas the SVM is a non-probabilistic binary classifier that favours fewer classes, k-NN is an instance-based algorithm that uses the similarity measures of the audio features to find the best match for a given new instance; while Random Forest is an ensemble algorithm that leverages the desirable potentials of 'weaker' models for better predictions. We compare the discrimination abilities of the classifiers using both individual domain feature set and combined domain feature set. The classification process involves the following steps:

## 4.1 Feature Selection

Best of the discriminatory audio features were selected using two attribute-selection algorithms namely – Correlation Feature Selection (CFS) and Principal Components Analysis (PCA). The original feature set consists of 12 attributes as highlighted in Table 1. However, the best first three features selected by CFS were varRMS, stdZCR and varSB; while the highest-ranking features according to PCA were meanRMS, arm-RMS, meanSF, stdZCR and varSF. It is interesting to note that the three features selected by CFS were good representation of the audio properties we considered earlier in the study. Whereas varRMS provides information on the energy level of the audio signal, stdZCR shows the periodicity, while varSB represents the spread or flatness of the audio spectral shape in terms of frequency localization.

## 4.2 Training and Testing

The experimental processes – STFT, Feature Extraction and Classification were carried out on Android Studio 1.5.1 Integrated Development Environment (IDE). With embedded Weka APIs, the classifier models were programmatically trained on the mobile devices running on Android 4.2.2 and 5.1.1, which were also used to record some of the audios used to evaluate the performance of the algorithms in real-time. The classification model was built for recognition and discriminating of respiratory signals with related sound features. We opted to train the models directly on the mobile devices rather than porting desktop-trained models, due to serialization and compatibility issues with android devices. Moreover, the response time of building the model on the smartphone is faster compared to the performance on the desktop. The machine learning algorithms are trained by using the statistical window-level features obtained from the audio signal frames. Due to limited datasets, a 'leave-one-out' strategy for 10-fold cross validation was used in the training and evaluation of the performance of the classifiers and the selected features. Statistical metrics used in the performance evaluation were precision, recall and F-measure.

## 5 Results

Here, we discuss the results and performance of the machine learning algorithms using different criteria. We also evaluated the real-time performance of the mobile device benchmarked on the CPU and memory usage as well as execution time of each of the modules in the entire process.

### 5.1 Performance of the Classifiers

In the evaluation of the classification process, we presented different scenarios of the problem to the classifiers in order to understand the mechanisms of their performances. The criteria used are as follows:

**Different Categories of Datasets.** We used two categories of datasets – 2.5 seconds length and 5 seconds length of the audio symptoms. The 2.5s length dataset has a total of 163 records (Wheeze = 49, Stridor = 33, Cough = 27, Clear-Throat = 26, Other = 28), while the 5s dataset used in the classification consists of 99 instances in total. Though there were fewer instances in the 5s datasets, the algorithms performed better on this category than in 2.5s datasets. This implies that longer audio durations rather than the number of instances provided the classifiers with more information to learn about the audio patterns.

**Scaling the number of classes and features.** Scaling the number of classes and features used in the classification also had much impact on the performance of the classifiers. From Table 2, we observed that the SVM classifier performed much better when we reduce the number of symptom classes to two; however, reducing only the number of features decreased the performance significantly. The k-NN algorithm on the other hand, performed fairly well in all the scenarios but showcased its best with reduced number of features and classes. Likewise, RF maintain its robustness with notable improvement in the performance when the number of features are few.

**Table 2.** Overall accuracy of the classifiers with scaled number of classes and features

| Classifiers | All Features & All Classes **m =12**, **n =5** | Reduced # of features **m =3** | Reduced # of classes **n =2** | Reduced # of features & classes **m=3**, **n =2** |
|---|---|---|---|---|
| k-NN | 0.88 | 0.88 | 0.89 | 0.92 |
| SVM | 0.75 | 0.59 | 0.80 | 0.78 |
| RF | 0.86 | 0.91 | 0.87 | 0.89 |

**Adjustment of the algorithms' parameters.** By increasing the complexity parameter $C$ of the SVM from 1.0 to 3.0, the classifier performance improved by 4.6%. However, setting the parameter $k$ of the k-NN Classifier to 1, gives an accuracy of 88.88 % but drops to 53.98% when $k$ is set to 5.

To further evaluate the discriminatory ability of the classification features, we examined two groups of classes whose elements are often conflated given the high level of their resemblance. These are: *Wheeze vs. Stridor* and *Cough vs. Clear Throat*. According to medical experts, these respiratory sounds are very common in exercise-induced VCD and bronchoconstriction or bronchial asthma. The comparisons are shown in Fig. 2 and Fig. 3 respectively. Fig. 2 indicates that though wheeze and stridor signals relatively have uniform oscillation (periodicity), stridor has a 'flatter' spectral shape given its wide frequency range. We also noticed that the classifiers generally found it difficult differentiating between cough and throat clearing. However, when presented with only time-domain features, the discrimination became clearer as shown in Fig. 3.

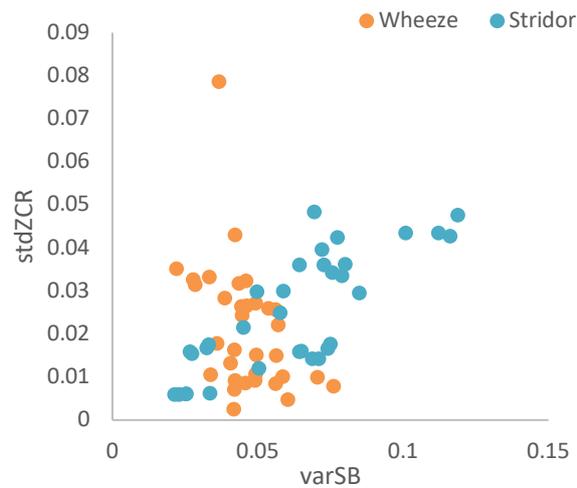

**Fig. 2.** Discriminating ability of time-frequency domain features – stdZCR and varSB on wheeze and stridor [17].

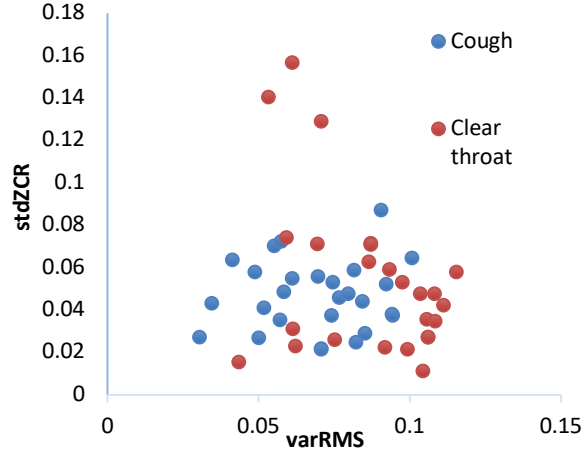

**Fig. 3.** Discrimination of cough from throat clearing by time-domain features – stdZCR and varRMS [17].

Though we experimented with three classifiers, we settled for only one - Random Forest, given its robustness in different scenarios as show in Table 3. The classifier also has a reasonable response time on the device and thus, was used for the real-time implementation of the classification tool.

**Table 3.** Weighted average performance for all classes with the three best CFS features

|  | Precision | Recall | F-measure |
|---|---|---|---|
| k-NN | 0.889 | 0.889 | 0.887 |
| SVM | 0.668 | 0.596 | 0.585 |
| RF | 0.918 | 0.919 | 0.918 |

As we were unable to get real-time access to clinical respiratory sound symptoms such as wheezes and stridor at the time of writing this paper; we performed an experimental test on the discriminatory ability of the classification tool in real-time, using records of common sound symptoms – cough and clear throat volunteered by asymptomatic individuals and those with pathological conditions. Fig. 4 shows correctly detected cough and clear- throat sounds in real-time on an android phone (Huawei p6 Ascend). The classification tool was also able to predict correctly, offline recorded wheeze and stridor sounds (Figs. 5a & 5b). By mere visualization, we can observe that the waveforms and the spectrograms of these sounds are different from each other. This may as well serve as a clue to physicians in the differential diagnosis of the underlying respiratory illnesses.

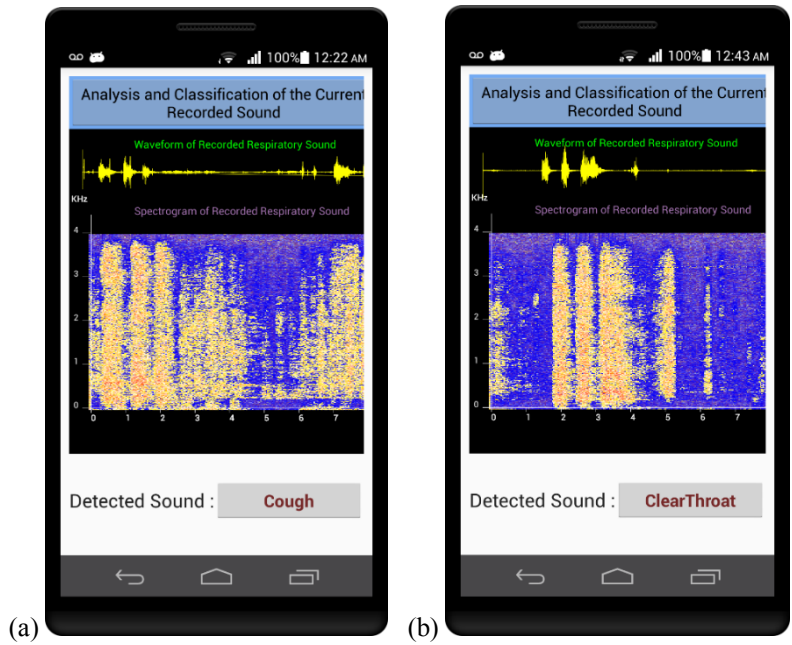

(a)                  (b)

**Fig. 4 (a) and (b).** Detected cough and clear-throat sounds in real-time.

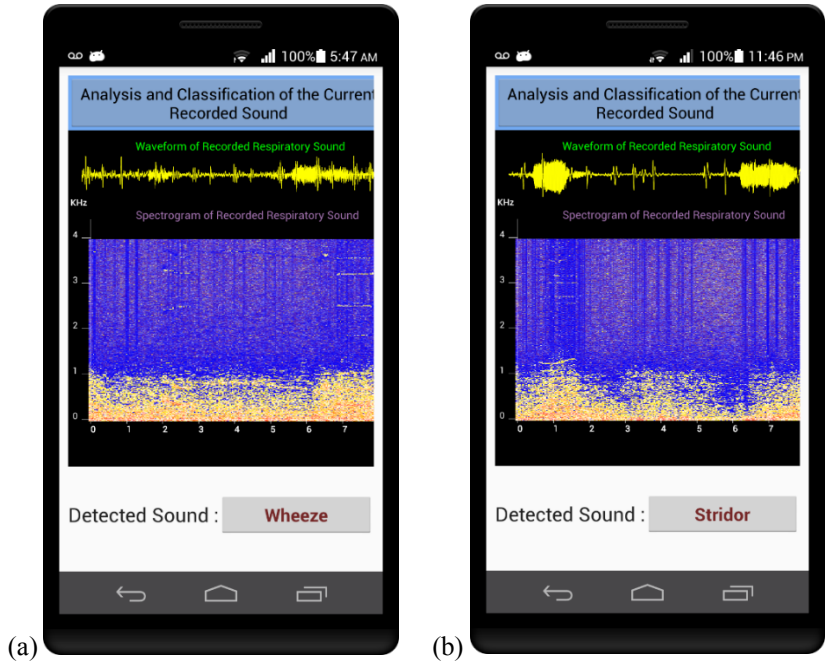

(a)                  (b)

**Fig. 5 (a) and (b).** Detected wheeze and stridor sounds recorded offline.

## 5.2 Device Performance on Resource Usage

We evaluate the smartphone performance on the utilization of the system resources when executing the major modules of the machine learning in real-time. The modules include audio pre-processing (framing and FFT), feature extraction, and the classification. Table 4 shows the measurements on the consumption of the device resources during the application run-time. The execution time in milliseconds (ms) is profiled in the android code. As expected, the response time for the pre-processing module was a bit long due to FFT metrics which are numerically intensive on the resources.

**Table 4.** Benchmarks on device resource usage by the major operations [17]

| Module | CPU | Memory | Response Time |
|---|---|---|---|
| Pre-processing | 27% | 2.2MB | 1404 ms |
| Feature Extraction | 25% | 8MB | 556 ms |
| Classification | 0.02% | 2MB | 722 ms |

# 6 Application

The application of the results obtained from the machine learning can be demonstrated using the following case scenarios in the study domain; where the mobile phone is used to monitor exercise-induced respiratory conditions (EIRCs) e.g. asthma, bronchospasm, rhinitis, and VCD.

## 6.1 Symptom Tracking

One of the important applications of this study is in assisting patients to keep track of the sound symptoms of their conditions. The study provides a visualized summary of the captured events on daily basis. For instance, Fig. 6 shows a column chart that measures the frequency of each sound symptom occurrence at specific periods of the day.

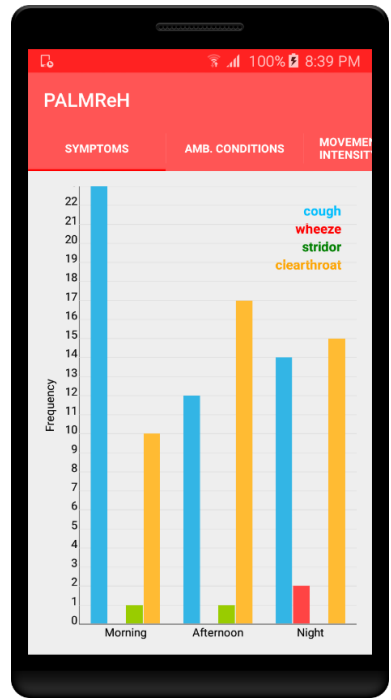

**Fig. 6.** Daily summary of the frequency of respiratory sound symptoms.

## 6.2 Integrated real-time monitoring of sound symptoms and contextual data for patient's self-management

| ID | Sound_Detected | Activity_Level | Relative_Humidity | Temperature_°C | Event_Time | Date |
|---|---|---|---|---|---|---|
| 31 | null | null | 80.33 | 28.44 | 04:49:48 | Apr 08 2017 |
| 32 | Cough | Vigorous | 80.76 | 28.34 | 04:49:53 | Apr 08 2017 |
| 33 | Cough | Vigorous | 80.76 | 28.12 | 04:49:58 | Apr 08 2017 |
| 34 | Cough | Vigorous | 81.55 | 27.83 | 04:50:03 | Apr 08 2017 |
| 35 | Cough | Vigorous | 82.78 | 27.33 | 04:50:08 | Apr 08 2017 |
| 36 | Cough | Moderate | 84.02 | 26.93 | 04:50:13 | Apr 08 2017 |
| 37 | null | null | 84.57 | 26.95 | 04:51:37 | Apr 08 2017 |
| 38 | Wheeze | Vigorous | 84.58 | 27 | 04:51:42 | Apr 08 2017 |
| 39 | Wheeze | Vigorous | 84.96 | 26.75 | 04:51:47 | Apr 08 2017 |
| 40 | Cough | Vigorous | 85.81 | 26.45 | 04:51:52 | Apr 08 2017 |

| PREVIOUS | NEXT |
|---|---|

Error Messages will be displayed here

**Fig. 7.** Embedded SQLite database in the smartphone that captures sound symptoms and contextual evidences of EIRCs.

Fig. 7 shows an embedded database (SQLite) that automatically captures both the symptoms and the contextual evidences by the smartphone. This aspect of the study also affords the user real-time access and instantaneous knowledge of the monitored events which are displayed graphically as illustrated in Fig.6 (the chart). For instance, the most frequently detected sound symptom, the level of physical activity, as well as the variations in the ambient temperature and humidity within the monitoring period. Using this knowledge, the user can correlate the captured events based on the context. For example, knowing the specific period of the day (e.g. morning, noon or night) when a symptom gets worse and the triggers that aggravate the symptoms, would help the patient to personally manage and control his/her respiratory condition. The generated information can also be processed into a report which can be stored on the mobile device or shared with healthcare providers and physicians for subsequent actions.

## 7 Conclusions

The difficulty in differentiating related respiratory sound symptoms at times, leads to subjective evaluation in the medical assessment of a patient which may result in misdiagnosis and undertreatment of the associated ailments. Though many researchers have proposed alternative approaches for objective detection and classification of respiratory sound symptoms using computer-based systems; a few of these approaches have been successfully performed exclusively on mobile phones. Leveraging the improvement on the storage and computational capabilities of modern mobile phones, we advanced the use of smartphone to detect and classify these sounds in real-time scenarios. And the tools we employed here were machine learning algorithms and standard sets of both temporal and spectral features of the audio signals often used in vocal and lung sound analysis. The study recorded over 83% accuracy on the average, in the classification process. We also illustrated the practical applications of the results in the study domain. We believe the information obtained from the process can aid physicians in further diagnosis of the suspected respiratory conditions; and also assist patients in the control and management of their conditions.